**WhatsApp security and role of metadata in preserving privacy**

Nidhi Rastogi, James Hendler
Rensselaer Polytechnic Institute, Troy, NY, USA
raston@rpi.edu
hendler@cs.rpi.edu

**Abstract:** WhatsApp messenger is arguably the most popular mobile app available on all smart-phones. Over one billion people worldwide for free messaging, calling, and media sharing use it. In April 2016, WhatsApp switched to a default end-to-end encrypted service. This means that all messages (SMS), phone calls, videos, audios, and any other form of information exchanged cannot be read by any unauthorized entity since WhatsApp version 2.16.2 (released April 2016). In this paper we analyze the WhatsApp messaging platform and critique its security architecture along with a focus on its privacy preservation mechanisms. We report that the Signal Protocol, which forms the basis of WhatsApp end-to-end encryption, does offer protection against forward secrecy, and MITM to a large extent. Finally, we argue that simply encrypting the end-to-end channel cannot preserve privacy. The metadata can reveal just enough information to show connections between people, their patterns, and personal information.

This paper elaborates on the security architecture of WhatsApp and performs an analysis on the various protocols used. This enlightens us on the status quo of the app security and what further measures can be used to fill existing gaps without compromising the usability. We start by describing the following (i) important concepts that need to be understood to properly understand security, (ii) the security architecture, (iii) security evaluation, (iv) followed by a summary of our work. Some of the important concepts that we cover in this paper before evaluating the architecture are - end-to-end encryption (E2EE), signal protocol, and curve25519. The description of the security architecture covers key management, end-to-end encryption in WhatsApp, Authentication Mechanism, Message Exchange, and finally the security evaluation. We then cover importance of metadata and role it plays in conserving privacy with respect to whatsapp.

**Keywords**: WhatsApp, privacy, security, Facebook, signal protocol, curve25519

## 1. Introduction

WhatsApp messenger was started by two ex-Yahoo employees (Business Insider 2015) and was sold to Facebook in 2014(WhatsApp Blog – Facebook 2016) but remained operationally independent. Since then, the user base has increased tremendously and over a billion users per day now use the app. As of January 2016, the average number of daily messages exchanged over WhatsApp is reported to be an astounding 34 billion (The Verge 2014). WhatsApp has been able to attract this unprecedented success because of its availability on all popular mobile operating systems, and is free of cost (or costs a nominal $0.99 per year). Free calls, unlimited messages, and media exchange, along with an easy to operate interface make it favorable for novice users as well.

However, as far as security is concerned, WhatsApp has come under fire several times in the past. The negligence shown towards making the application secure made it an easy target for attackers. For example, in 2011, a problem was found in the app verification process proving that the authentication mechanism was unsecure (Schrittwieser et. al 2012). Researchers were able to exploit valid usage session by successfully hijacking several user accounts (called session hijacking). This allowed unauthorized access where an attacker could spoof the sender identification, thus receiving messages targeted to the victim. A packet sniffer could then intercept the traffic and log all communication details. All later attempts were either a half-baked attempt to encrypt messages or were broken at launch. This lax approach continued and by the time it was may 2012, WhatsApp was still sending messages in plain text, which means there was no encryption for any kind of communication.

In the wake of increasing privacy concerns and the war between Apple and FBI over encryption of phone data, WhatsApp has switched to end-to-end encryption. This has enabled the messenger app user to send all communication encrypted. It is no more easy for an unauthorized person to read text messages, videos, audios, or files by surreptitiously listening to the communication as data is no more send in plaintext.

This paper elaborates on the security architecture of WhatsApp and analyzes the various protocols used. We perform an extensive literature study from several online resources on Whatsapp and related concepts and use that to understand the working of the application and its security protocols. Also, while whatsapp is a popular app for the mobile platform, its computer version can be accessed via a web browser or by installing an app for the windows or mac OS platform. Since a phone number is required as the primary identification of a user, the QR code needs to be scanned to authorize the computer (WhatsApp FAQ – WhatsApp Web).

We also take a closer look at the app security and what further measures can make it stronger without compromising usability. In the following sections, we cover some important security concepts applicable to WhatsApp, understand and evaluate the security architecture, measures taken to ensure user privacy, make recommendations on improvements, and finally end with a summary of our work.

## 2. Security Fundamentals
### 2.1 End-to-End Encryption (E2EE)
This is a system of communication, which allows only the communicating parties to access the messages because the medium is encrypted. In theory, no eavesdropper can access the cryptographic keys needed to decrypt the conversation. This includes service providers like cellular companies, ISPs, and app developers. Theoretically, an adversary cannot access the transmitted data even after the traffic has been intercepted. This is possible because of the various properties of the encryption protocols used for making the end-to-end communication encrypted and inaccessible for an unauthorized user. In the figure below, the communication channel between the two phones or computers is encrypted.

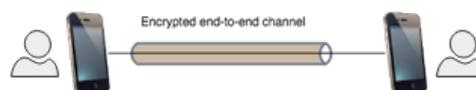

Figure 1 - E2E encryption between two smartphones.

### 2.2 Signal Protocol
Signal Protocol (previously Axolotl) enables end-to-end encryption in WhatsApp. It is used to encrypt both text messages and voice calls by using an asynchronous method under a shared key. The protocol was chosen as it can provide plausible deniability and forward-secret asynchronous communications, among other features, on mobile devices. (Praetorian 2015)

### 2.3 Plausible deniability
By deniability or repudiation, it means that a message receiver can be sure where the message originated from but cannot prove the identity of the sender. In essence, the sender can deny being the person who originally sent the message (Open Whisper Systems 2013). Signal protocol uses a compact derivative of the Off-the-Record (OTR) protocol to enable this feature. Before we get into any further details, let's first understand the working of the signal protocol.

Each member participant in a WhatsApp conversation has a long-term identity key that they use to sign an ephemeral key. This ephemeral key is exchanged among members to calculate a shared secret, typically using Diffie–Hellman (D-H) key exchange method. D-H allows the participants to jointly establish a shared secret key, which can then be used to encrypt subsequent communications.

The shared secret from this key exchange is used to derive three keys for each participant - a sending and a receiving cipher key, and a set of MAC keys. These MAC keys confirm message authenticity and integrity, and are included in every transmitted message. Notice here that the MAC keys are subsequently derived from the original shared key ensuring that the message was indeed sent by the claimed sender. At the same time, both the parties are involved in generating the shared key as well as the subsequent MAC keys (also called ephemeral keys). While this keeps the message integrity intact, the authenticity of sending the message that they originated can be denied later. This is because of the shared key, which makes the receiver capable of producing a sender's MAC key.

**2.4 Forward secrecy**

If the encryption keys from a user's smartphone or computer somehow get compromised, a fresh key for every new message is issued. This prevents an adversary from not only deriving the ephemeral keys but also from using it to decrypt any message transmitted in the past.

Signal Protocol uses the following types of keys:
1. Identity key pair, a long-term Curve25519 key pair generated at install time for all asymmetric cryptographic operations.
2. Signed pre Key, a medium term Curve25519 key pair.
3. Pre Keys, also Curve25519 keys but for one-time use. These are used to actually encrypt the message.

Signal Protocol uses a compact derivative of OTR where it uses D-H exchange in each key generation step above, which continually ratchets the key material forward. This is the underlying principle behind forward secrecy as the keys that finally encrypt the message are ephemeral. Recording the encrypted traffic cannot divulge the key material or decrypt previous messages. Even if a device is physically compromised, no keys at any given time are stored on the device that can help an adversary decrypt previously exchanged ciphertext. Note that this property is very different from the traditional ways of encrypting data in motion or at rest. In these cases, the same key or a periodically changed key (which is usually a slow process) is used to encrypt data. This makes it extremely important to store the key at a secure location, lest all the recorded messages ever exchanged, and sometimes with all different parties, may get into the hands of the adversary. By contrast, the key exchange mechanism in signal protocol is ephemeral. Hence, if a key is ever compromised in the future, all recorded ciphertext will remain private.

There are other advantages for choosing signal protocol. It is a mobile-friendly end-to-end (e2e) protocol, which can decrease the size of packets by using protobufs. Protobuf, or protocol buffer, is a small logical record of information, containing a series of name-value pair that offer an automated mechanism for serializing structured data. It works similar to XML but differs by being faster, smaller, and simpler.

**2.5 Curve25519**

Elliptic-curve based cryptographic (ECC) systems are public-key cryptosystems that rely on the inability to determine n from Y = nX, where X and Y are publicly known base points.
Curve25519 helps compute the public part of this equation, which is 128-bits in length.
Curve25519 is also an ECC curve, which is a variant of the Diffie-Hellman protocol. For this reason, Curve25519 can be successfully implemented with the elliptic curve Diffie–Hellman (ECDH) key agreement scheme. This property enables Curve25519 to compute shared keys that can be exchanged over unencrypted channels as well. As mentioned in earlier sections, each member on WhatsApp has a long-term identity key that is used to calculate this shared secret.

Curve25519 introduced by Daniel J. Bernstein (Bernstein 2006), computes very fast in terms of key compression, key validation, and timing-attack protection among others. This makes the curve a practical choice for large-scale implementation, as is the case with WhatsApp.

**3. Security Architecture**

Now that we understand some key concepts, we can turn our attention to the implementation and key exchanges that take place in WhatsApp. See figure 2 for more details.

3.1 **Key Management**

*3.1.1. End-to-End Encryption working in WhatsApp*

Each WhatsApp user possesses a long-term key that is stored on the device memory, not readily accessible to the user. This key is used to create another shared key using which a WhatsApp user can securely communicate with another use. A secure communication channel is established between the two, and it remains intact until events such as app reinstall, device change, etc. The following steps describe key management in the flow diagram shown in figure 2. The initiating client is called initiator, and the requesting client is called the recipient.

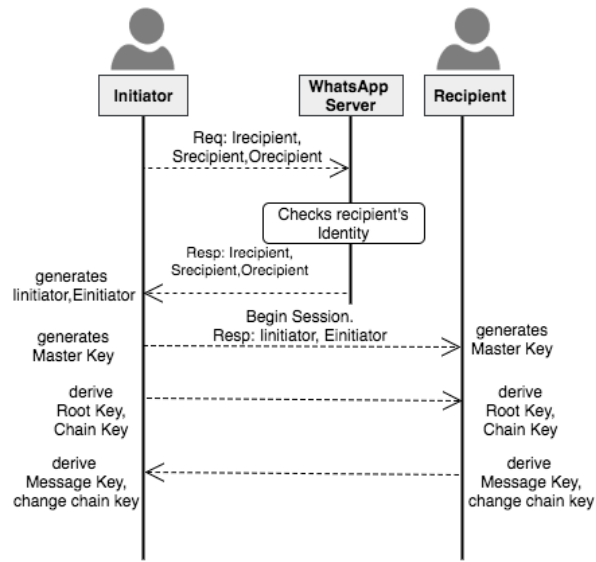

Figure 2: Flow diagram of whatsapp end-to-end encryption

1. The initiator requests the public identity key, public signed pre key, and a single public one-time pre key for the recipient. The identity key, called **Irecipient** is a long-term curve22519 key pair. The signed pre key, called **Srecipient** is a medium-term curve22519 key pair, and signed by **Irecipient**. The one-time pre-key, called **Orecipient** is a list of curve22519 key pairs mainly for one time use. All these keys are generated during installation, reinstallation, or change of device.
2. The WhatsApp server returns the requested public key values to the initiating client. The one-time pre key is ephemeral, and remains on the server only until requested.

3. The initiator saves the keys requested in step 1 and generates an ephemeral curve25519 key pair, called **Einitiator**, and loads its own identity key, called **Iinitiator**.
4. Using these keys generated and requested in the above step the initiator can now calculate a shared secret with the recipient -
Master Key - ECDH(Iinitiator, Srecipient) || ECDH(Einitiator, Irecipient) || ECDH(Einitiator, Srecipient) || ECDH(Einitiator, Orecipient)
This master key is used to create subsequent session keys between the two parties. A Hashed Message Authentication Code (HMAC)-based key derivation function (HKDF) derives the root key, and chain keys from the master key. It takes master key as the input keying material and extracts from it a fixed-length pseudo random key. This key expands into several additional pseudorandom keys, resulting in the root and chain keys, both with 32-byte value.
5. The server contacts the recipient using the member id lookup and sends session information with the initiator - Einitiator and Iinitiator.
6. Using this session information, the recipient calculates at its end shared secret, which is the master key and confirms integrity of the message and has been sent unaltered by an authorised person. Recipient also deletes the ephemeral, one-time pre-key, **Orecipient**.
7. Using the chain keys generated in previous steps, parties involved in the conversation generate a message key of 80-byte value. This encrypts each message and ratchets forward the chain key used to derive the message key, every time a message is sent in a given session. This works by increasing a counter that is part of a function deriving the chain key. This is a key step in providing forward secrecy, as the chain key is no more of use for messages sent earlier and hence cannot be used to decrypt them. With the chain key changing with every message, the message key also changes having a similar effect on forward message encryption.

Message key = HMAC-SHA256(chain key, 0x01)
Chain key = HMAC-SHA256(chain key, 0x02)

WhatsApp also uses QR code verification method for out-of-band user verification. The QR code contains, among other things, a 32-byte Irecipient and Iinitiator - which are the public identity keys for both users. Another way to get a similar experience is by comparing a 60-digit number.

**4. Security and Privacy Evaluation**
Signal Protocol drastically reduces the possibility of having a man-in-the-middle attack. This is primarily because OTR is based on a mechanism where it uses D-H exchange in each key generation step mentioned earlier. This continually ratchets the key material forward. For an active adversary who has managed to decrypt the channel, the integrity of the encryption keys can to be traced all the way back to the original shared key, which requires a fair amount of time and key tracking. One can be assured that no MITM attack is possible on any of the subsequently generated keys.

However, a major security concern is worth mentioning here. While WhatsApp messages are secure in transit, most of the endpoint devices – such as smartphones, tablets, and computers – do not encrypt the data residing on them in the same way that Apple does with its most recent iPhone. WhatsApp offers to backup messages likely on a cloud server. Some of the options given are Google drive, Apple iCloud, etc. We do not have any information about message encryption on the cloud platform yet, unless WhatsApp decides to share these details soon.

Also, WhatsApp does not offer encryption of past communication at app level, which can expose the user messages in case of device theft.

**4.1 Privacy implication of plausible deniability**
The prevalence of global surveillance has caused much concern to many users. Some of the concerns have been related to a third party listening to user conversations, without permission. Another one is being held against a message they sent in the past in the court of law.

Signal protocol was designed keeping such privacy concerns in mind, among other security issues. For this purpose, it ensures that the message sender or receiver cannot be irrefutably tied to a particular message sent in the past by using the various ratchet forward encryption techniques.

In the privacy domain, there have been concerns related to user metadata as well. WhatsApp encrypts the communication channel between users using end-to-end encryption. The metadata of the user is encrypted as well when data is in motion on the communication channel between various parties. It is essential to understand that information stored in metadata is just as important in preserving privacy of the users, as is the data itself. The company's legal terms allow them to store information associated with successfully delivered messages such as time of delivery, mobile phone numbers involved in the messages, size of any digital content swapped between the two parties (Bernstein 2006). Also, the app persists the user to share one's entire contact list with the app. This is a way to further gather information about who is in a particular social network of a user. It is like trading the convenience of having the app to figure out who uses it amongst one's contacts for giving up the entire list of which one contacts regularly, including those who don't use the app. There is still no option of selectively adding contacts to the WhatsApp list. Any addition of this feature in the future will not help existing users as they have already shared this detail with the app.

A smartphone metadata reflects a wealth of details both at the level of individual calls and when analyzed in aggregate. Computer scientists and researchers have proved this a number of times in the past. It is here where WhatsApp falters. While the metadata is encrypted during transit, phone numbers, timestamps, connection duration, connection frequency, as well as user location are being stored on the company's servers. This metadata is sufficient to create a profile and draw some strong inferences between the communicating parties. And as we've seen very often, both governments and hackers can get their hands on the metadata if they really go after it.

What advantage would Facebook, the parent company has in addition to the metadata related information coming via WhatsApp? WhatsApp had vowed that it would not be selling advertisements. However, there is no condition that can stop its parent company from doing so by using information gathered through the whatsapp. In combination to one's activities on Facebook, it can potentially help create a more accurate understanding of the user behavior, and social interactions thereby serving as a strong measure of profiling for some targeted ads. This is not truly a major concern as long as the user sees ads that make sense to them. Any change in the content delivery algorithm can lead to a very different user experience, where in some cases the user may outright stop using the app.

For group chat, the communication initiator sends message to the whatsapp server, which in turn distributes it to all the group members. This is a very easy way of for Facebook to learn all about ones social interactions and communities. A lot can be deduced by performing some kind of traffic analysis just by using the metadata like from the message volume exchanged.

In August 2016, WhatsApp changed its terms of privacy where it stated that it plans to transfer user data to its parent company, Facebook. It had earlier promised that this data would not be disclosed or used for marketing purposes. But now it will share user account information with Facebook and the Facebook family of companies, like the phone number the user used as a primary identifier. The companies intend to use WhatsApp account information to show users "more relevant ads on Facebook" and to send users marketing messages via WhatsApp. A phone number is like a digital social security number (EPIC - WhatsApp). It can uniquely identify a person as this information is provided every time when filling up forms for various purposes. It can also connect various sources of data, like health records, financial data, and education, online presence, etc. and create a full profile of a person.

Metadata can also provide enough information about the user who relies on the platform provider to deliver content. This content can sometimes lead to influencing their opinion, for example political opinions. During the US presidential campaigns taking place in 2016, advertisements, videos, or posts reached out to a fairly wide audience. The coverage provided by Facebook is unparalleled in comparison to the coverage provided by any other platform. Ones that focus too much on a certain negative or positive aspect of republican candidate Donald Trump or democrat candidate, Hillary Clinton can lead a user to create a bias view of the candidate over a period of time.

**5. Conclusion**
WhatsApp has attracted a lot of attention because of its large-scale usage of end-to-end encryption, a first of its kind. The purpose it to re-instate users trust in using chat apps without worrying about privacy and security concerns. In this paper, we went over the various fundamental of advanced cryptography protocols that enable the various security and privacy properties of whatsapp. We discussed these features and how successful whatsapp has been by deploying them in their security architecture. We also went over the privacy concerns that still remain due to metadata remaining unencrypted and within the territory of the app provider.